\documentclass[aps,twocolumn,pra,superscriptaddress,articlepaper,nofootinbib,longbibliography]{revtex4-2}
\usepackage{amsmath, amsthm, amssymb, amsfonts}
\usepackage{physics}
\usepackage{mathtools}
\usepackage{upgreek}
\usepackage{booktabs}
\usepackage{url}
\usepackage{float}
\usepackage{enumitem}
\AtBeginDocument{
\heavyrulewidth=.08em
\lightrulewidth=.05em
\cmidrulewidth=.03em
\belowrulesep=.65ex
\belowbottomsep=0pt
\aboverulesep=.4ex
\abovetopsep=0pt
\cmidrulesep=\doublerulesep
\cmidrulekern=.5em
\defaultaddspace=.5em
}

\usepackage{txfonts}
\usepackage[pdftex]{graphicx}
\usepackage[usenames, dvipsnames, svgnames, table]{xcolor}
\usepackage[colorlinks=true, citecolor=Blue,linkcolor=BrickRed, urlcolor=magenta]{hyperref}
\usepackage{bm}
\usepackage{placeins} % To use \FloatBarrier to output all floats which are not yet typeset
\usepackage{setspace}
\usepackage[final]{changes}
\usepackage{comment}

\usepackage{soul} % cancel line on texts \st

\usepackage{cleveref}
\crefname{section}{Section}{Sections}
\crefname{appendix}{Appendix}{Appendices}
\crefname{figure}{Figure}{Figures}
\crefname{table}{Table}{Tables}
\crefname{equation}{Eq.}{Eqs.}
\crefname{assumption}{Assumption}{Assumptions}

\usepackage{siunitx}

\newcommand{\OSI}[2]{$\mathcal{O}\left(\SI{#1}{#2}\right)$}

\usepackage{dsfont} % operators
\usepackage{mathtools}
\usepackage{empheq}
\newcommand{\mrm}[1]{\mathrm{#1}}

\newcommand{\prtHz}{\per\sqrt{\mrm\Hz}}

\begin{document}
% \title{On the application of optical frequency comb to gravitational-wave detection in space:\\ experimental study and its practicality}
% \title{Alternative approach to time-delay interferometry with optical frequency comb:\\ toy model and experimental demonstration}
\title{Alternative approach to time-delay interferometry with optical frequency comb}

\author{Kohei Yamamoto}
\email{y9m9k0h@gmail.com}
\affiliation{Center for Space Sciences and Technology, University of Maryland, Baltimore County, 1000 Hilltop Circle, Baltimore, MD 21250, USA}
\affiliation{NASA Goddard Space Flight Center, 8800 Greenbelt Road, Greenbelt, MD 20771, USA}
\affiliation{Center for Research and Exploration in Space Science and Technology, NASA/GSFC, 8800 Greenbelt Road, Greenbelt, MD 20771, USA}

\author{Hannah Tomio}
\affiliation{Department of Aeronautics and Astronautics, Massachusetts Institute of Technology, 77 Massachusetts Ave., Cambridge, MA 02139, USA}

\author{Charlotte Zehnder}
\affiliation{Department of Physics, University of Arizona, 1118 E. Fourth Street, Tucson, AZ 85721, USA}

\author{Kenji Numata}
% \affiliation{Lasers and Electro-optics Branch, NASA/GSFC, 8800 Greenbelt Road, Greenbelt, MD 20771, USA}
\affiliation{NASA Goddard Space Flight Center, 8800 Greenbelt Road, Greenbelt, MD 20771, USA}

\author{Holly Leopardi}
\affiliation{NASA Goddard Space Flight Center, 8800 Greenbelt Road, Greenbelt, MD 20771, USA}

\begin{abstract}
Spaceborne gravitational wave observatories, exemplified by the Laser Interferometer Space Antenna (LISA) mission, are designed to remove laser noise and clock noise from interferometric phase measurements in postprocessing.
The planned observatories will utilize electro-optic modulators (EOMs) to encode the onboard clock timing onto the beam phase.
Recent research has demonstrated the advantage of introducing an optical frequency comb (OFC) in the metrology system with the modified framework of time-delay interferometry (TDI): the removal of the EOM and the simultaneous suppression of the stochastic jitter of the laser and the clock in the observation band.
In this paper, we explore an alternative approach with the OFC-based metrology system.
We report that after proper treatment, it is possible to use the measured carrier-carrier heterodyne frequencies to monitor the time derivative of the pseudoranges, which represent the physical light travel time and the clock difference.
This approach does not require changing the existing TDI framework, as previous OFC based efforts did.
Furthermore, this approach naturally captures not only stochastic jitter but also clock offsets and slow drifts.
We also present the experimental demonstration of our scheme using two separate systems to model two spacecraft.
Using this novel approach, we synchronize the two independent phase measurement systems with an accuracy better than \SI{0.47}{\nano\second}, while the stochastic jitter in the observation band is suppressed down to the setup sensitivity around the LISA performance levels at \SI{15}{\pico\meter\prtHz}.
\end{abstract}

\maketitle

\section{Introduction}\label{sec:intro}
Since the first detection of a gravitational wave (GW) from stellar-mass binary black holes in 2015 by the Laser Interferometer Gravitational Wave Observatory (LIGO) detectors~\cite{GW150914}, the field has been rapidly advancing.
Today, GW observations by ground-based detectors are no longer a rare event~\cite{LIGO:Catalog1,LIGO:Catalog2,LIGO:Catalog3}.
To open a new observation channel in the millihertz band, international research teams have studied and proposed several spaceborne GW detector concepts.
One of these spaceborne observatories, the Laser Interferometer Space Antenna (LISA) mission led by the European Space Agency (ESA)~\cite{LisaRed}, is planned to be launched in 2035.

LISA forms a nearly equilateral triangle constellation with three spacecraft separated over \SI{2.5}{million\,\kilo\meter}.
The three spacecraft exchange laser beams to configure interspacecraft interferometers.
GW signals are accumulated over the interspacecraft laser links and encoded in the phase of the laser beams, which are measured by an onboard phasemeter.
Given realistic LISA orbits provided by ESA~\cite{LisaOrbits}, the relative positions of the spacecraft drift by around $10^8$\,\si{\meter} with a relative speed of up to \SI{10}{\meter\per\second}.
This results in the spacecraft constellation having unequal-arm interferometer geometries.
The resulting phase signal is overwhelmed by laser frequency noise, masking the \SI{10}{\pico\meter}-level fluctuations imparted by the GW signals.
To suppress the laser noise by around 8 orders of magnitude, a postprocessing technique called time-delay interferometry (TDI) will be utilized to carefully time-shift the phase signals from the three spacecraft with \si{nanosecond} accuracy to virtually synthesize equal-arm interferometers.
This in turn requires proper treatment of the onboard clock differences between the three spacecraft~\cite{Heinzel:Ranging}.
For this purpose, each spacecraft in the LISA constellation embeds its onboard clock information on the outgoing beam phase via an electro-optic modulator (EOM), and monitors the clock difference via the phase of a heterodyne beatnote between its own clock modulation sideband and that of the incoming beam~\cite{Hartwig2021,Yamamoto2022}.
The measured clock differences can be used to remove the clock jitter in the observation band and to correct the slow drift of the timestamps in postprocessing.

% Tinto and Yu~\cite{Tinto2015} have analytically shown that the laser and clock noise can be suppressed by means of an optical frequency comb (OFC)~\cite{Jones2000,Udem2002,Fortier2019}, which would remove the need for the EOMs and clock modulation of the interspacecraft link.
Tinto and Yu~\cite{Tinto2015} have analytically shown that the laser and clock noise can be suppressed by means of an optical frequency comb (OFC)~\cite{Jones2000,Udem2002,Fortier2019}, which has been generalized to given TDI combinations later by Tan et al.~\cite{Tan2022}.
This would remove the need for the EOMs and clock modulation of the interspacecraft link.
By using the laser carrier as a reference to lock one of the OFC modes, a radio frequency (RF) signal that is coherent with the reference laser can be generated.
If we use the OFC-generated RF signal as a system clock for the onboard phase measurement systems, the laser noise and the clock noise are no longer independent in the phasemeter output data streams.
As a result, the adaptation of TDI with scaling factors for the phase signals enables the suppression of these noise factors simultaneously.
This concept has been experimentally demonstrated by simulating the large delay between spacecraft using acousto-optic modulators with a single phase measurement system~\cite{Vinckier2020}.

In this paper, we propose an alternative noise suppression approach based on the use of OFCs.
This approach leverages the carrier-carrier beatnote frequency not only for the main scientific purpose of detecting GWs but also for monitoring the so-called \emph{pseudorange} (or, to be precise, its time derivative): the difference between the time of the light emission in the time frame of the onboard clock on an emitting spacecraft and the time of the light reception in the time frame of the onboard clock on a receiving spacecraft.
%which encompasses the clock difference and the light travel time between spacecraft.
% Compared with LISA, OFCs enable us to push the role of the clock sideband to the beam carrier with the GW signals left intact in TDI combinations.
% Since this approach follows that of LISA except for the method of extracting the pseudoranges, the traditional TDI framework can be applied with no modification.
Since this approach follows that of LISA except for the method of extracting the pseudoranges, it preserves the conventional TDI formalism.
In addition, our approach does not detract from the original advantage mentioned in~\cite{Tinto2015}: namely, the simultaneous suppression of the laser and clock noises, as the conventional TDI framework can directly work on the pseudoranges and suppress the laser and clock noise in one step as well~\cite{Hartwig2022}.
Furthermore, while in the original OFC-based approach, the discussion was limited to the stochastic jitter in the GW observation band, this alternative approach also extracts information of the clock frequency offset and drift from the carrier-carrier beatnotes.
The results presented here maximize the technological advantages of incorporating OFCs.
Finally, we describe an experimental demonstration using two separate systems that model two spacecraft to study the ability to extract the total clock difference.

\section{Theoretical description}\label{sec:theory}
\begin{figure}[h]
    \centering
    \includegraphics[width=8.6cm]{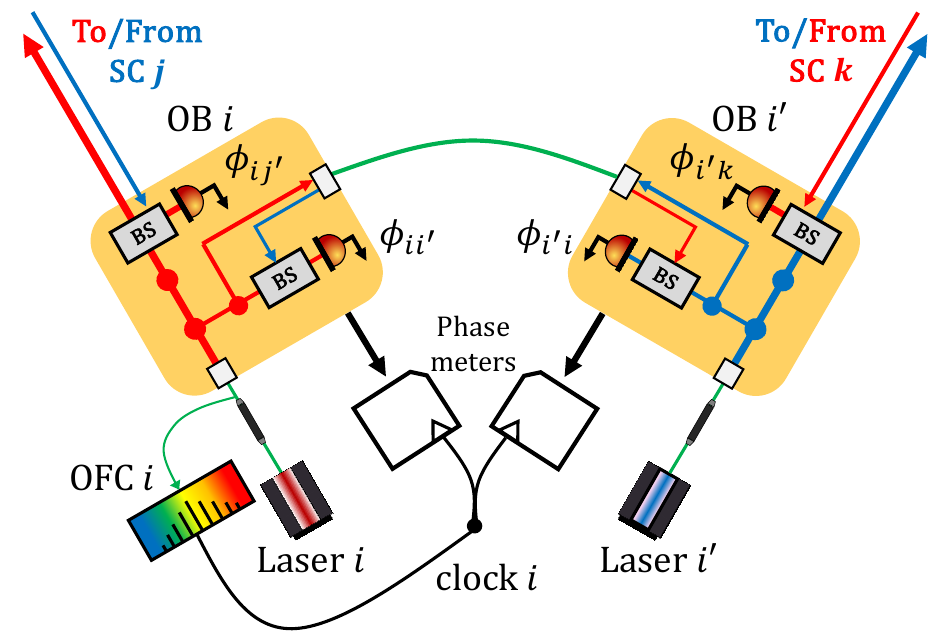}
    \caption{
    Conceptual diagram of a possible spacecraft payload equipped with an OFC, labeled by $i$ while having two remote spacecraft $j$ and $k$.
    The system on the left, considered the \emph{primary}, is labeled after the spacecraft, while the system on the right, considered the \emph{secondary}, is labeled with the primed index.
    The primary and secondary lasers are drawn in red and blue, respectively.
    Optical interference occurs at each beam splitter.
    Only the interspacecraft and reference interferometers are depicted, neglecting the test-mass interferometers.
    A spacecraft onboard clock can be generated by the OFC locked to the primary laser $i$.
    SC: spacecraft; OB: optical bench; BS: beam splitter; OFC: optical frequency comb.}
    \label{fig:payload}
\end{figure}

We begin by providing a theoretical description of our scheme, focusing on how to extract the time derivative of the interspacecraft pseudorange from the carrier-carrier beatnotes.
After proper treatment, these beatnotes can serve the same role as the clock sideband-sideband beatnotes in the LISA-like EOM-based approach.
Our model is constructed based on a possible payload equipped with an OFC, as depicted conceptually in \cref{fig:payload}.
In addition to the interspacecraft interferometers beating a local beam with a received beam from a distant spacecraft, we also use so-called reference interferometers that primarily monitor the frequency difference between the two local lasers.
Although the actual payload also needs test mass interferometers to monitor the test mass motion relative to the optical bench, these are omitted in the figure as they are not necessary to derive the pseudorange.
Regarding the notation, the three spacecraft are labeled by $i,j,k\in{1,2,3}$.
As shown in \cref{fig:payload}, the \emph{primary} system, on the left, generates an onboard clock via an OFC and is labeled after the spacecraft, while the associated \emph{secondary} system, on the right and in the same spacecraft, is labeled by its primed index, namely $i'$, $j'$, or $k'$.
While we derive the time derivative of the pseudorange via the carrier-carrier beatnote, a complete pseudorange also includes a large time offset, which can be derived using pseudorandom noise ranging~\cite{Sutton2010,Esteban2011,Xie2023,Yamamoto2024,Euringer2024,Yamamoto2025}.
However, this is out of the scope of this paper, and we assume throughout this work that the complete pseudorange can be derived in postprocessing, for example, via TDI-ranging (TDIR)~\cite{Tinto:TDIR}.
A short notation summary is given in~\cref{tab:notation}.

\begin{table}[b]
    \caption{\label{tab:notation}
    Notation summary.
    A quantity without a superscript (or with the superscript $\tau_m$ in \cref{sub:onboard_frame}) is considered according to spacecraft proper time.
    In contrast, a quantity with a superscript $\tau_i$ ($i\in\{i,j,k,1,2,3\}$) is expressed according to the time frame of onboard clock $i$.
    }
    \begin{ruledtabular}
    \begin{tabular}{ccc}
     & In proper time & In clock time $i$\\
    \hline \\\vspace{1.5mm}
    Frequency of laser $j$ & $\nu_j$ or $\nu_j^{\tau_m}$ & $\nu_j^{\tau_i}$ \\\vspace{1.5mm}
    OFC-related frequencies & $f_x$ or $f_x^{\tau_m}$ & $f_x^{\tau_i}$\\\vspace{1.5mm}
    Clock time $j$ & $\tau_j$ or $\tau_j^{\tau_m}$ & $\tau_j^{\tau_i}$\\
    \end{tabular}
    \end{ruledtabular}
\end{table}

\subsection{Lasers and clocks}\label{sub:laser_clock}
Let us start with a model of the frequency of a laser and an OFC-based clock.
All signals in this subsection are written according to the proper time of the spacecraft.

We model the laser frequency on spacecraft $i$ with the following three terms:
\begin{align}
    \nu_i(\tau) &= \nu_c + O_i(\tau) + \dot{p}_i(\tau).
    \label{eq:nu_i}
\end{align}
where each term has the following definition and order of magnitude: the nominal central frequency of the onboard laser $\nu_c\sim\SI{281.6}{\tera\Hz}$, the large offset frequency $O_i\sim$ \OSI{10}{\mega\Hz} assuming a LISA-like heterodyne bandwidth, and the stochastic jitter $\dot{p}_i\sim\SI{100}{\Hz\prtHz}$ in the observation band (letting $p_i$ be the laser phase noise following convention).

We generate a clock signal from the laser frequency $\nu_i$ using an OFC.
In general, the frequency of a given mode of OFC\,$i$ with integer mode number $n_i$ is written as:
\begin{align}
    f_{n_i} &= n_if_{\mrm{rep},i} + f_{\mrm{ceo},i},
    \label{eq:fn}
\end{align}
where $f_{\mrm{rep},i}$ is the repetition rate and $f_{\mrm{ceo},i}$ is the carrier-envelope offset frequency of the OFC.
To have $f_{\mrm{rep},i}$ coherent with the laser frequency, two control loops must be properly locked: the $f_\mrm{ceo}$ loop that detects the carrier-envelope offset via the so-called ``$f-2f$" self-referencing scheme~\cite{Jones2000,Telle1999} and that actuates the pump power of the comb laser, and an offset-lock loop on the heterodyne beatnote between a reference laser $\nu_\mrm{i}$ and the comb mode $f_{n_i}$, which actuates the cavity length of the comb laser.
Given the carrier-envelope offset is locked to a setpoint $f_{\mrm{ceo},i}$, the offset-lock loop robustly achieves the following:
\begin{align}
    f_{\mrm{OPT},i} &= \nu_i(\tau) - f_{n_i}(\tau),
    \label{eq:fOPTi}
\end{align}
where $f_{\mrm{OPT},i}$ is the set point of the offset-lock loop.
Note that $f_{\mrm{OPT},i}$ in \cref{eq:fOPTi} can be either positive or negative depending on the relation between $\nu_i$ and $f_{n_i}$.
Solving \cref{eq:fOPTi} for $f_{\mrm{rep},i}$ results in:
\begin{align}
    f_{\mrm{rep},i}(\tau) &= \frac{\nu_i(\tau) - f_{\mrm{OPT},i} - f_{\mrm{ceo},i}}{n_i}
    \nonumber\\
    &= \frac{\nu_c}{n_i}\left(1 + \frac{O_i(\tau) + \dot{p}_i(\tau) - f_{\mrm{OPT},i} - f_{\mrm{ceo},i}}{\nu_c}\right).
    \label{eq:fOFC_i2}
\end{align}
This is the onboard clock frequency written in the frame of the spacecraft proper time.

If we define a nominal clock frequency in this frame as $f_c$, our onboard clock has the following clock error in units of fractional frequency:
\begin{align}
    \dot{q}_i &= \frac{f_{\mrm{rep},i}(\tau)-f_c}{f_c}
    \nonumber\\
    &= \frac{\nu_c}{n_if_c}\left(1 + \frac{O_i(\tau) + \dot{p}_i(\tau) - f_{\mrm{OPT},i} - f_{\mrm{ceo},i}}{\nu_c}\right) - 1
    \nonumber\\
    &= \frac{n_c}{n_i}\left(1 + \frac{O_i(\tau) + \dot{p}_i(\tau) - f_{\mrm{OPT},i} - f_{\mrm{ceo},i}}{\nu_c}\right) - 1,
    \label{eq:dotqi}
\end{align}
under
\begin{align}
    f_c &\coloneqq \nu_c/n_c,
    \label{eq:fc}
\end{align}
where $n_c$ is the nominal mode number that represents a certain comb mode.
One practical way to determine the nominal values is the following: (1) define the nominal clock rate $f_c$ to be consistent with the sampling rate used in the data processing; (2) set $\nu_c$ at a reasonable middle frequency in the laser frequency range; (3) set $n_c = [\nu_c/f_c]$; (4) fine-tune $\nu_c$ to meet $\nu_c = n_cf_c$.

\subsection{Onboard clock frame}\label{sub:onboard_frame}
We then transform the reference time frame from the proper time of the spacecraft to that of the onboard clock.
This gives the actual onboard measurements.

In general, a clock time $\tau_i$ as a function of another time $\tau_m$, acting as a reference frame, can be expressed by:
\begin{align}
    \tau_i^{\tau_m}(\tau) &= \tau + \delta\tau_i^{\tau_m}(\tau),
    \label{eq:taut}\\
    \delta\tau_i^{\tau_m}(\tau) &= q_i^{\tau_m}(\tau) + T_{0,i}^{\tau_m},
    \label{eq:deltaut}
\end{align}
where we call $\delta\tau_i^{\tau_m}(\tau)$ the \emph{timer deviation}, which is composed of the clock deviation $q_i^{\tau_m}(\tau)$ and an initial offset of the clock start time $T_{0,i}^{\tau_m}$.
The argument $\tau$ is a given time counted by the reference frame: $\tau_i^{\tau_i}(\tau) = \tau$.
Inversely, $\tau_m$ can be expressed according to $\tau_i$ as
\begin{align}
    \tau_m^{\tau_i}(\tau) &= \tau + \delta\tau_m^{\tau_i}(\tau) = \tau - \delta\tau_i^{\tau_m}\left(\tau_m^{\tau_i}(\tau)\right),
    \label{eq:taum}
\end{align}
which can be acquired by substituting $\tau_m^{\tau_i}$ into $\tau$ in \cref{eq:taut} under the identity $\tau^{\tau_i}_m\left(\tau^{\tau_m}_i(\tau)\right)=\tau$.

The next step is to relate the phase of an electromagnetic wave in different time frames.
As a scalar quantity, the phase $\phi$ of an electromagnetic wave can transform from one time frame to another using a simple time shift~\cite{HartwigPhD}:
\begin{align}
    \phi^{\tau_i}(\tau) &= \phi^{\tau_m}\left(\tau_m^{\tau_i}(\tau)\right)
    \nonumber\\
    &= \phi^{\tau_m}\left(\tau + \delta\tau_m^{\tau_i}(\tau)\right).
    \label{eq:phi_i2m}
\end{align}
Hence, the frequency of the electromagnetic wave $\nu$ can be derived via the time derivative of \cref{eq:phi_i2m} as:
\begin{align}
    \nu^{\tau_i}(\tau) &= \frac{d\phi^{\tau_i}(\tau)}{d\tau}
    \nonumber\\
    &= \nu^{\tau_m}\left(\tau_m^{\tau_i}(\tau)\right)\cdot\frac{d\tau_m^{\tau_i}(\tau)}{d\tau}
    \nonumber\\
    &= \frac{\nu^{\tau_m}\left(\tau_m^{\tau_i}(\tau)\right)}{1+\dot{q}_i^{\tau_m}\left(\tau_m^{\tau_i}(\tau)\right)}.
    \label{eq:nu_tau_i}
\end{align}

We apply this general expression to our laser and clock models in \cref{eq:nu_i,eq:dotqi}. In the following, $\tau_m$ represents the proper time of the spacecraft, $\tau_i$ represents the OFC-generated onboard clock, and $\nu$ is replaced by the laser frequency $\nu_i$.
Substituting \cref{eq:nu_i,eq:dotqi} into \cref{eq:nu_tau_i}, the frequency of laser $i$ according to the onboard clock $\tau_i$ is:
\begin{align}
    \nu^{\tau_i}_i(\tau) &= \frac{n_i}{n_c}\cdot\frac{\nu_c + O_i\left(\tau_m^{\tau_i}(\tau)\right) + \dot{p}_i\left(\tau_m^{\tau_i}(\tau)\right)}{1 + \frac{O_i\left(\tau_m^{\tau_i}(\tau)\right) + \dot{p}_i\left(\tau_m^{\tau_i}(\tau)\right) - f_{\mrm{OPT},i} - f_{\mrm{ceo},i}}{\nu_c}}
    \nonumber\\
    &\approx \frac{n_i}{n_c}\cdot \left(\nu_c  + f_{\mrm{OPT},i} + f_{\mrm{ceo},i}\right)\cdot \left(1 - \frac{O_i(O_i-f_{\mrm{OPT},i}-f_{\mrm{ceo},i})}{\nu_c^2}\right)
    \nonumber\\
    &\approx \frac{n_i}{n_c}\cdot \left(\nu_c  + f_{\mrm{OPT},i} + f_{\mrm{ceo},i}\right).
    \label{eq:nutaui}
\end{align}
It is reasonable, in practice, to assume that $f_{\mrm{OPT},i}$, $f_{\mrm{ceo},i}$, and $O_i$ all have an order of magnitude of \SI{10}{\mega\Hz}.
Therefore, in the second line, the last term is around $(\SI{10}{\mega\Hz}/\SI{281}{\tera\Hz})^2\approx1.3\cdot10^{-15}$.
Compared with the synchronization accuracy of \SI{3.3}{\nano\second} required for LISA, this can be considered negligible over a reasonably long time of \SI{1000000}{\second}, which results in the last line.
\cref{eq:nutaui} suggests that the frequency of laser $i$ looks constant in the time frame of the onboard clock. This is expected because the OFC-generated clock coherently changes with the laser.

In a similar manner, we can derive the frequency of laser $i'$ in the local OFC clock frame as:
\begin{align}
    \nu^{\tau_i}_{i'}(\tau) &= \frac{n_i}{n_c}\cdot\frac{\nu_c + O_{i'}\left(\tau_m^{\tau_i}(\tau)\right) + \dot{p}_{i'}\left(\tau_m^{\tau_i}(\tau)\right)}{1 + \frac{O_i\left(\tau_m^{\tau_i}(\tau)\right) + \dot{p}_i\left(\tau_m^{\tau_i}(\tau)\right) - f_{\mrm{OPT},i} - f_{\mrm{ceo},i}}{\nu_c}}
    \nonumber\\
    &\approx \frac{n_i}{n_c}\cdot \left(\nu_c + O_{ii'}\left(\tau_m^{\tau_i}(\tau)\right) + \dot{p}_{ii'}\left(\tau_m^{\tau_i}(\tau)\right) + f_{\mrm{OPT},i} + f_{\mrm{ceo},i}\right),
    \label{eq:nutaui'}
\end{align}
where we define a given quantity with two subscripts $x_{ij}$ as 
\begin{align}
    x_{ij}(\tau) \coloneqq x_j(\tau) - x_i(\tau).
    \label{eq:xij}
\end{align}
We use this notation throughout the rest of this paper.

From \cref{eq:nutaui,eq:nutaui'} the frequency measurement of the reference interferometer, i.e., the beatnote between the lasers $i$ and $i'$, is given as:
\begin{align}
    \nu^{\tau_i}_{ii'}(\tau) &= \nu^{\tau_i}_{i'}(\tau) - \nu^{\tau_i}_{i}(\tau)
    \nonumber\\
    &= \frac{n_i}{n_c}\cdot \left(O_{ii'}\left(\tau_m^{\tau_i}(\tau)\right) + \dot{p}_{ii'}\left(\tau_m^{\tau_i}(\tau)\right)\right).
    \label{eq:nuii'_taui}
\end{align}

\subsection{Pseudorange estimation}\label{sub:pseudorange}

Finally, we show how to estimate the time derivative of the interspacecraft pseudorange using the carrier-carrier beatnote measurements.
The pseudorange is defined as the time difference between the emission and the reception of the laser light according to the onboard clocks on the emitting and receiving spacecraft, respectively~\cite{Hartwig2022}.
% Hence, the phase of the laser $j$, namely $\phi_j$, at the reception by spacecraft $i$ in the time frame of its clock $i$ is related via the pseudorange $d^{\tau_i}_{ij}$ to the phase of the laser at its emission by spacecraft $j$ in the time frame of the clock $j$ as follows:
Hence, the phase of the laser $j$, namely $\phi_j$, at the reception by spacecraft $i$ according to clock $i$ is related via the pseudorange $d^{\tau_i}_{ij}$ to $\phi_j$ at its emission by spacecraft $j$ according to clock $j$ as follows:
\begin{align}
    \phi^{\tau_i}_j(\tau) &= \phi^{\tau_j}_j\left(\tau - d^{\tau_i}_{ij}(\tau)\right).
    \label{eq:phij_taui}
\end{align}

Ideally, we measure the phase of a beatnote between the primary lasers associated with the onboard clocks.
Let us assume that this measurement exists for now; in that case, such an interspacecraft interferometer at spacecraft $i$ reads:
\begin{align}
    \phi^{\tau_i}_{ij}(\tau) &= \phi^{\tau_i}_j(\tau) - \phi^{\tau_i}_i(\tau)
    \nonumber\\
    &= \phi^{\tau_j}_j\left(\tau - d^{\tau_i}_{ij}(\tau)\right) - \phi^{\tau_i}_i(\tau).
    \label{eq:phiij_taui}
\end{align}
The measurement in units of frequency $\nu^{\tau_i}_{ij}$ can be derived by the time derivative,
\begin{align}
    \nu^{\tau_i}_{ij}(\tau) &= \frac{d\phi^{\tau_i}_{ij}(\tau)}{d\tau}
    \nonumber\\
    &= \dot{\mathbf{D}}^{\tau_i}_{ij}\nu^{\tau_j}_j(\tau) - \nu^{\tau_i}_i(\tau),
    \label{eq:nuij_taui}\\
    \dot{\mathbf{D}}^{\tau_i}_{ij}f(\tau) &\coloneqq \left(1 - \dot{d}^{\tau_i}_{ij}(\tau)\right)\cdot f\left(\tau - d^{\tau_i}_{ij}(\tau)\right).
    \label{eq:dotDij_taui}
\end{align}
\cref{eq:dotDij_taui} defines the Doppler delay operator with an arbitrary function $f(\tau)$, following the notation in~\cite{Hartwig2022}.

Each frequency is expressed in the time frame of the associated onboard clock.
Therefore, substituting \cref{eq:nutaui} into \cref{eq:nuij_taui} gives
\begin{align}
    \nu^{\tau_i}_{ij}(\tau) &= \frac{d\phi^{\tau_i}_{ij}(\tau)}{d\tau}
    \nonumber\\
    &= \left(1 - \dot{d}^{\tau_i}_{ij}(\tau)\right)\cdot\frac{n_j}{n_c}\left(\nu_c  + f_{\mrm{OPT},j} + f_{\mrm{ceo},j}\right) 
    \nonumber\\
    &\hspace{10mm}- \frac{n_i}{n_c}\left(\nu_c  + f_{\mrm{OPT},i} + f_{\mrm{ceo},i}\right).
    \label{eq:nuij_taui2}
\end{align}
Solving this equation for the time derivative of the pseudorange $\dot{d}^{\tau_i}_{ij}$, results in:
\begin{align}
    \dot{d}^{\tau_i}_{ij}(\tau) &= 1 - \frac{n_i}{n_j}\cdot\left(1 + \frac{\frac{n_c}{n_i}\nu^{\tau_i}_{ij}(\tau) - f_{\mrm{OPT},ij} - f_{\mrm{ceo},ij}}{\nu_c}\right),
    \label{eq:dotdij_taui}
\end{align}
where $f_{\mrm{OPT},ij}=f_{\mrm{OPT},j}-f_{\mrm{OPT},i}$ and $f_{\mrm{ceo},ij}=f_{\mrm{ceo},j}-f_{\mrm{ceo},i}$ are the differences in the set point of the OFC feedback loops between the two spacecraft.

This derivation is based on the assumption that we have a direct measure of $\nu^{\tau_i}_{ij}$. 
However, this is not the case in a real system because the primary laser $i$ faces the secondary laser $j'$ on the opposing spacecraft; that is, only a direct measurement of $\nu^{\tau_i}_{ij'}$ is possible.
Likewise, with regard to the pseudorange time derivative $\dot{d}^{\tau_i}_{ik}$, the secondary laser $i'$ is linked to the primary laser $k$ of the other opposing spacecraft, resulting in the beatnote $\nu^{\tau_i}_{i'k}$.
Ultimately, we need to combine the interspacecraft interferometer measurements with those of the reference interferometers, given in \cref{eq:nuii'_taui}, to accurately estimate the time derivative of the pseudoranges.

First, following \cref{eq:dotdij_taui}, let us naively estimate the left- and right-hand pseudoranges using the actual interspacecraft interferometers as:
\begin{subequations}\label{eq:hatdotd_leftright_taui}
    \begin{align}
        \hat{\dot{d}}^{\tau_i, (0)}_{ij}(\tau) &= 1 - \frac{n_i}{n_j}\cdot\left(1 + \frac{\frac{n_c}{n_i}\nu^{\tau_i}_{ij'}(\tau) - f_{\mrm{OPT},ij} - f_{\mrm{ceo},ij}}{\nu_c}\right),
        \label{eq:hatdotdi'j_taui}\\
        \hat{\dot{d}}^{\tau_i, (0)}_{ik}(\tau) &= 1 - \frac{n_i}{n_k}\cdot\left(1 + \frac{\frac{n_c}{n_i}\nu^{\tau_i}_{i'k}(\tau) - f_{\mrm{OPT},ik} - f_{\mrm{ceo},ik}}{\nu_c}\right),
        \label{eq:hatdotdi'k_taui}
    \end{align}
\end{subequations}
where the interspacecraft interferometers are
\begin{subequations}\label{eq:nu_leftright_taui}
    \begin{align}
        \nu^{\tau_i}_{ij'}(\tau) &= \dot{\mathbf{D}}^{\tau_i}_{ij}\nu^{\tau_j}_{j'}(\tau) - \nu^{\tau_i}_i(\tau),
        \label{eq:nuij'_taui}\\
        \nu^{\tau_i}_{i'k}(\tau) &= \dot{\mathbf{D}}^{\tau_i}_{ik}\nu^{\tau_k}_{k}(\tau) - \nu^{\tau_i}_{i'}(\tau).
        \label{eq:nui'k_taui}
    \end{align}
\end{subequations}
The hat symbol in \cref{eq:hatdotd_leftright_taui} indicates that this is an estimate, rather than a true value.
Let us call \cref{eq:hatdotd_leftright_taui} the zeroth estimate, denoted by the integer superscripts $(0)$.

The zeroth estimates of the pseudoranges in \cref{eq:hatdotd_leftright_taui} are expected to lack sufficient accuracy because the secondary lasers with the primed indices are not directly related to the clocks.
However, ideally, we can derive the virtual beatnote frequency between the primary lasers by combining the interspacecraft and reference interferometers as:
\begin{subequations}\label{eq:nuideal_taui}
    \begin{align}
        \nu^{\tau_i}_{ij}(\tau) &= \nu^{\tau_i}_{ij'}(\tau) - \dot{\mathbf{D}}^{\tau_i}_{ij}\nu^{\tau_j}_{jj'}(\tau),
        \label{eq:nuij_taui_D}\\
        \nu^{\tau_i}_{ik}(\tau) &= \nu^{\tau_i}_{i'k}(\tau) + \nu^{\tau_k}_{ii'}(\tau).
        \label{eq:nuik_taui}
    \end{align}
\end{subequations}
However, the Doppler delay operator in \cref{eq:nuij_taui_D} itself includes the pseudorange $d^{\tau_i}_{ij}$ and its time derivative $\dot{d}^{\tau_i}_{ij}$, which we seek to estimate.
Therefore, we can iterate over the following self-consistent equations to achieve an estimate with high accuracy:
\begin{subequations}\label{eq:bardotd_leftright_taui}
    \begin{align}
        \hat{\dot{d}}^{\tau_i,(m+1)}_{ij}(\tau) &= \hat{\dot{d}}^{\tau_i,(0)}_{ij}(\tau) + \frac{n_c}{n_j}\frac{1}{\nu_c}\hat{\dot{\mathbf{D}}}^{\tau_i,(m)}_{ij}\nu^{\tau_j}_{jj'}(\tau),
        \label{eq:bardotdi'j_taui}\\
        \hat{\dot{d}}^{\tau_i,(1)}_{ik}(\tau) &= \hat{\dot{d}}^{\tau_i,(0)}_{ik}(\tau) - \frac{n_c}{n_k}\frac{1}{\nu_c}\nu^{\tau_i}_{ii'}(\tau),
        \label{eq:bardotdi'k_taui}
    \end{align}
\end{subequations}
where $m$ is an integer representing the iteration order.

The right-hand pseudorange $\dot{d}^{\tau_i}_{ik}$ in \cref{eq:bardotdi'k_taui} results in the true value with a single correction (that is, $\hat{\dot{d}}^{\tau_i,(1)}_{ik}=\dot{d}^{\tau_i}_{ik}$).
The left-hand pseudorange in \cref{eq:bardotdi'j_taui} is less straightforward due to the Doppler delay operator.
Concerning this iterative operation, let us represent the error of the $m$-th estimate of the left-hand pseudorange $\hat{\dot{d}}^{\tau_i,(m)}_{ij}$ by $\hat{\dot{\delta}}^{\tau_i,(m)}_{ij}$, where:
\begin{align}
    \hat{\dot{\delta}}^{\tau_i,(m)}_{ij}(\tau) &= \hat{\dot{d}}^{\tau_i,(m)}_{ij}(\tau) - \dot{d}^{\tau_i}_{ij}(\tau).
    \label{eq:hatdotdeltaij}
\end{align}
In the following, we assume that the integrated version $\hat{\delta}^{\tau_i,(m)}_{ij}$ does not gain an additional error from the start-time offset, and $\hat{\delta}^{\tau_i,(m)}_{ij}$ can be easily calculated by the integral of \cref{eq:hatdotdeltaij}.

Taking into account \cref{eq:hatdotdeltaij} and \cref{eq:bardotdi'j_taui}, the $(m+1)$ error becomes
\begin{align}
    \hat{\dot{\delta}}^{\tau_i,(m+1)}_{ij}(\tau) &\approx -\frac{n_c}{n_j}\frac{1}{\nu_c}\left(\hat{\dot{\delta}}^{\tau_i,(m)}_{ij}\mathbf{D}^{\tau_i}_{ij}\nu^{\tau_j}_{jj'} + \hat{\delta}^{\tau_i,(m)}_{ij}\dot{\mathbf{D}}^{\tau_i}_{ij}\dot{\nu}^{\tau_j}_{jj'}\right)
    \nonumber\\
    &\approx -\frac{n_c}{n_j}\frac{1}{\nu_c}\cdot\hat{\dot{\delta}}^{\tau_i,(m)}_{ij}\mathbf{D}^{\tau_i}_{ij}\nu^{\tau_j}_{jj'}
    \nonumber\\
    &\sim O\left(\frac{\SI{10}{\mega\Hz}}{\SI{281}{\tera\Hz}}\cdot\hat{\dot{\delta}}^{\tau_i,(m)}_{ij}\right)
    \label{eq:hatdotdeltaij_mp1}
\end{align}
where we neglect a cross-term between $\hat{\dot{\delta}}^{\tau_i,(m)}_{ij}$ and $\hat{\delta}^{\tau_i,(m)}_{ij}$.
In the second line, the second term that scales with the time derivative of the heterodyne frequency is neglected, as the heterodyne frequency is expected to drift very slowly (less than \SI{10}{\Hz\per\second} at the fastest~\cite{LisaOrbits}).
Due to the suppression factor of $\SI{10}{\mega\Hz}/\SI{281}{\tera\Hz}$, we believe that the iteration rapidly converges on the true value, and a single iteration might be enough for most cases.
Note that it would also be possible to suppress the readout noise coupling by constructively combining the two reference interferometers in the same spacecraft, for example, with $\nu^{\tau_j}_{jj'}\rightarrow \left(\nu^{\tau_j}_{jj'} - \nu^{\tau_j}_{j'j}\right)/2$ in \cref{eq:bardotdi'j_taui}.

In analogy to the sideband-sideband beatnote in the case of LISA, this suggests that we can derive the time derivative of the pseudorange from the carrier-carrier beatnote measurements in the OFC-based metrology system, provided that we know the OFC mode numbers $n_i$ and $n_j$.
We provide a focused discussion on the resolution of the OFC mode number in \cref{app:resolve_n}.
% As long as the OFC mode numbers are identified and the pseudorange is derived accordingly, we can use the traditional TDI framework instead of modifying it as proposed in previous work ~\cite{Tinto2015,Tan2022}.
As long as the OFC mode numbers are identified and the pseudorange is derived accordingly, we can use the conventional TDI framework instead of modifying it as proposed in previous work ~\cite{Tinto2015,Tan2022}.

Note that the iterative operation with the reference interferometer in \cref{eq:bardotd_leftright_taui} is fairly analogous to the removal of right-hand modulation noise in the LISA-like EOM-based system~\cite{Hartwig2021}, which uses the sideband-sideband beatnotes in the reference interferometer.
The main differences are the following:
first, the modulation noise is only about the in-band jitter. In contrast, in the OFC case, all relative components of the laser frequencies (the offset, the slow drift, and the in-band jitter) need to be extracted through this iterative process;
second, the two reference interferometers in the same spacecraft are combined destructively to isolate the modulation noise from the laser noise in the EOM case, while they are combined constructively in the OFC case.

\section{Experiment}\label{sec:experiment}
We experimentally demonstrate the pseudorange extraction modeled in \cref{sec:theory} using two independent optical metrology systems (OMSs) that mimic the two spacecraft without the interspacecraft delay.
Under this condition, the pseudorange reduces to the pure clock difference, that is,
\begin{align}
    \dot{d}^{\tau_i}_{ij}(\tau) &\rightarrow - \dot{q}^{\tau_i}_j(\tau).
    \label{eq:dotd_dotq}
\end{align}
The minus sign is caused by the definition of the pseudorange in \cref{eq:phij_taui}: \emph{delay} is positive, meaning the slower clock error has a positive sign.
Note that $\dot{q}^{\tau_i}_j$ is the clock deviation of clock $j$ against the other clock $i$, unlike in \cref{sub:onboard_frame}, which defined the clock error against the proper time of the spacecraft. 
\cref{sub:setup} describes the experimental setup.
In \cref{sub:model}, we model the synchronization of the two systems in the experiment using the framework from \cref{sec:theory}, which is followed by the experimental results in \cref{sub:result}.

\subsection{Setup}\label{sub:setup}

\begin{figure*}
    \centering
    \includegraphics[width=17.2cm]{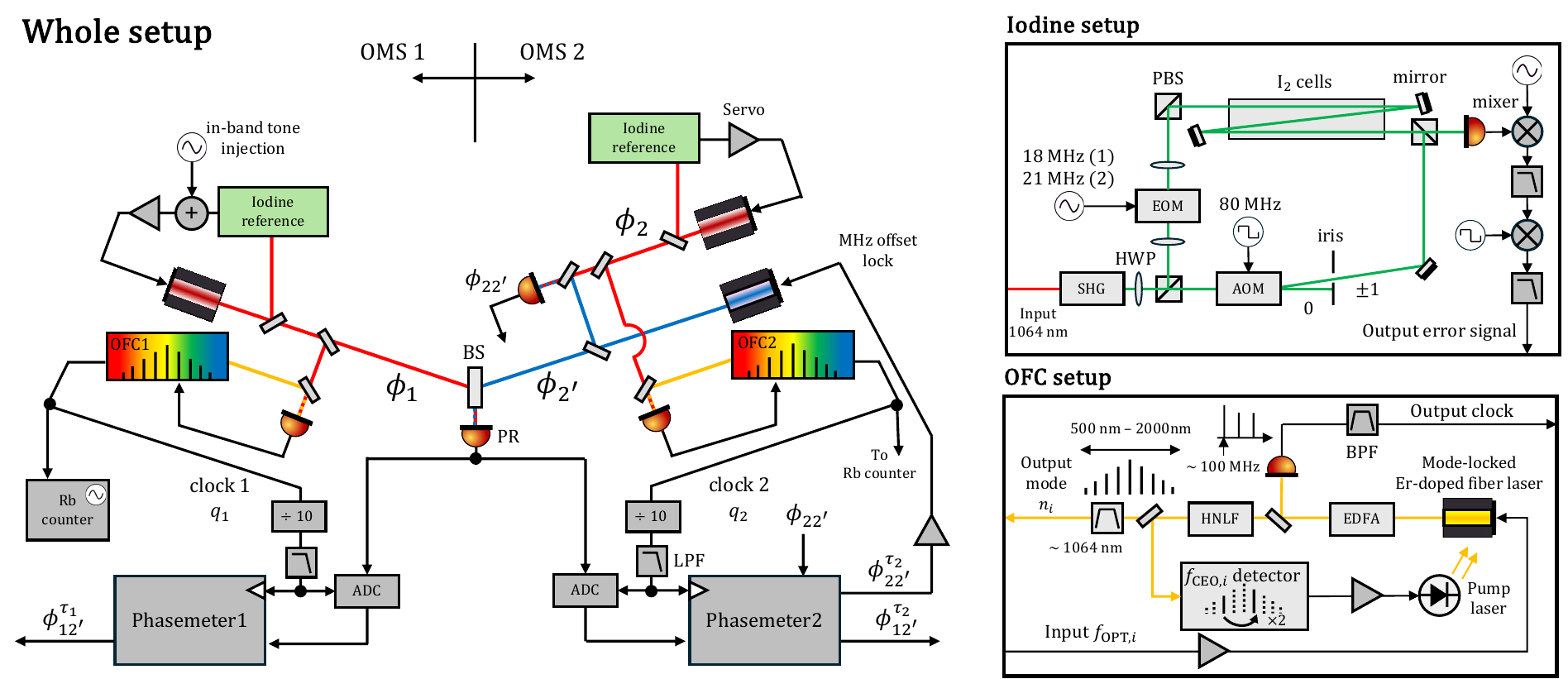}
    \caption{
    The experimental setup is composed of two optical metrology systems (OMSs), which mimic two spacecraft.
    The OFC setup and the free-space custom iodine setup are depicted on the bottom and top right inlays.
    The red and blue lasers are primary and secondary, as established in \cref{fig:payload}.
    The OFC is locked to the primary laser, $i$, which is stabilized with the custom iodine setup.
    The phase of the heterodyne beatnote between the primary laser from OMS\,1 and the secondary laser from OMS\,2 is individually extracted by the phasemeters driven by the independent OFC-generated clock signals in the OMSs.
    BS: beam splitter, ADC: analog-to-digital converter, PR: photo receiver, LPF: low-pass filter, BPF: band-pass filter, PBS: polarizing beam splitter, EOM: electro-optic modulator, AOM: acousto-optic modulator, SHG: second harmonic generator, EDFA: Er-doped fiber amplifier, HNLF: highly nonlinear fiber.}
    \label{fig:setup}
\end{figure*}

The experimental setup is depicted in \cref{fig:setup}.
This setup simulates two independent spacecraft using OMS\,1 and OMS\,2 on the left and right.
Each OMS is composed of a laser, an OFC, and a phasemeter.

The continuous wave (CW) laser in OMS\,1 is locked to an iodine absorption line using the pump-probe method~\cite{Hall1981}; see the top right in \cref{fig:setup}.
We choose Line 1110 (R(56)32-0) $a_1$~\cite{Ye1999}, which has the absolute doubled wavelength of \SI{1064.4911524}{\nano\meter} or, equivalently, \SI{281.629826}{\tera\Hz} in the absolute doubled frequency.
The absolute frequency reference, through the iodine cell, was introduced for an in-depth characterization of the system, as is described in \cref{sub:result}.
This stabilizes the laser frequency to \SIrange{100}{1000}{\Hz\prtHz} in the observation band.
In addition, to facilitate the characterization of the synchronization performance, a tone at \SI{1}{\Hz} was injected into the error signal of the stabilization loop.

In practice, there is a \SI{20}{\mega\Hz} offset in the infrared induced by an AOM driven at \SI{80}{\mega\Hz} for the pump.
The signs of the \SI{20}{\mega\Hz} offset are intentionally set opposite in OMS\,1 and OMS\,2 by selecting the opposite first-order diffraction beams, which enables the iodine-stabilized lasers to be easily characterized using their heterodyne beats.
As a result, the frequency difference between the two iodine-stabilized primary lasers is \SI{40}{\mega\Hz}.

A portion of the laser power is picked-off for OFC\,1.
The OFC module is depicted on the bottom right in \cref{fig:setup}.
The carrier-envelope offset frequency $f_{\mrm{ceo},i}$ is locked to a setpoint of \SI{6}{\mega\Hz} via the ``$f-2f$" method, while one comb mode is offset-locked to the primary laser with a setpoint of \SI{21}{\mega\Hz}.
As a result of those two control loops, the pulse repetition rate $f_{\mrm{rep},i}$ becomes coherent with the iodine-stabilized laser with a signification downconversion factor from the laser central frequency of $\sim\SI{281.6}{\tera\Hz}$ to the pulse repetition rate around $\SI{100}{\mega\Hz}$.
The optical pulse train is detected by a photoreceiver (PR) and bandpass filtered to extract a sinusoidal signal at \SI{100}{\mega\Hz}.
The Moku:Pro from Liquid Instruments is used as the phasemeter in both OMSs.
To use the OFC RF signal as the phasemeter clock, we convert the \SI{100}{\mega\Hz} signal to \SI{10}{\mega\Hz} using a divide-by-10 frequency divider, as is required by the Moku:Pro commercial hardware.

OMS\,2 is composed of the same components as OMS\,1, except that it also features the secondary laser (blue), which is offset-locked to the primary laser.
In this way, the two lasers in OMS\,2 play a similar role to those in the same spacecraft, as depicted in \cref{fig:payload}.
Hence, we can perform the pseudorange estimation including the reference interferometer, as modeled in \cref{eq:bardotdi'j_taui}, and by defining OMS\,2 as the remote spacecraft.
The offset lock between the primary and secondary lasers is set at \SI{28}{\mega\Hz}.
Because of the \SI{40}{\mega\Hz} offset between the primary lasers and the \SI{28}{\mega\Hz} offset between the primary and secondary lasers in OMS\,2, the heterodyne frequency for our ``interspacecraft" interferometer is \SI{12}{\mega\Hz}.
After converting the optical signal to an electrical signal by a PR, the signal is split and connected to the individual phasemeters in OMS\,1 and 2.
The phasemeters, driven by the associated OFC clocks, extract the heterodyne phase with phase-locked loops (PLLs), which are analyzed according to the model provided in the next subsection.

\subsection{Model}\label{sub:model}
Applying the general model in \cref{sec:theory} to our experiment, we show how to synchronize two OMSs in \cref{fig:setup}.
We assume that OMS\,1 is the primary module, and OMS\,2 is synchronized to OMS\,1.
In this case, the formulation of the left-hand pseudorange in \cref{eq:hatdotdi'j_taui} is applicable.
Hence, the time derivative of the clock deviation can be formulated, using the replacement in \cref{eq:dotd_dotq}, as
\begin{align}
    \hat{\dot{q}}^{\tau_1,(0)}_2(\tau) &= \frac{\hat{n}_1}{\hat{n}_2}\cdot\left(1 + \frac{\frac{n_c}{\hat{n}_1}\nu^{\tau_1}_{12'}(\tau) - f_{\mrm{OPT},12} - f_{\mrm{ceo},12}}{\nu_c}\right) - 1.
    \label{eq:dotq2_tau1}
\end{align}
In our experiment, we use the same set points for the OFC feedback loops: $f_{\mrm{OPT},1}=f_{\mrm{OPT},2}=\SI{21}{\mega\Hz}$ and $f_{\mrm{ceo},1}=f_{\mrm{ceo},2}=\SI{6}{\mega\Hz}$.
Therefore, $f_{\mrm{OPT},12}=f_{\mrm{ceo},12}=0$.
$\hat{n}$ is the fitted mode number, resolved as discussed in \cref{app:resolve_n}.

Following \cref{eq:bardotdi'j_taui}, we iterate the self-consistent equation, given below, until we reach sufficient accuracy, namely \SI{3.3}{\nano\second} targeted by LISA:
\begin{align}
    \hat{\dot{q}}^{\tau_1,(m+1)}_2(\tau) &= \hat{\dot{q}}^{\tau_1,(0)}_2(\tau) - \frac{n_c}{\hat{n}_2}\frac{1}{\nu_c}\hat{\dot{\mathbf{D}}}^{\tau_1,(m)}_{12}\nu^{\tau_2}_{22'}(\tau),
    \label{eq:bardotq2_tau1}
\end{align}

By integrating $\hat{\dot{q}}^{\tau_1,(m)}_2$ in time and adding a fitted start time offset $\hat{T}_{0,2}^{\tau_1}$, we get the timer deviation of clock $2$ against clock $1$, i.e., $i,m\rightarrow 2,1$ in \cref{eq:deltaut}, as
\begin{align}
    \delta\hat{\tau}_2^{\tau_1,(m)}(\tau) &= \int^\tau_0 \hat{\dot{q}}^{\tau_1,(m)}_2(\tau') d\tau' + \hat{T}_{0,2}^{\tau_1}.
    \label{eq:deltauj_taui}
\end{align}

Finally, we can combine the beatnote signals sampled by clock $i$ and $j$, namely $\phi^{\tau_1}_{12}(\tau)$ and $\phi^{\tau_2}_{12}(\tau)$, to suppress the laser and clock noise at the same time.
For this purpose, \cref{eq:phi_i2m} tells us that we can time-shift $\phi^{\tau_2}_{12}$ by $\delta\hat{\tau}_2^{\tau_1,(m)}$, and therefore the noise-free combination $\Delta^{\tau_1}_a$ reads
\begin{align}
    \Delta^{\tau_1,(m)}_a(\tau) &= \phi^{\tau_1}_{12}(\tau) - \phi^{\tau_2}_{12}\left(\tau + \delta\hat{\tau}_2^{\tau_1,(m)}(\tau)\right) \approx 0.
    \label{eq:Delta_taui_a}
\end{align}
Note that this does not require any scaling factor for the beatnote phase signals, unlike the original OFC-based approach described in~\cite{Tinto2015},
% which demonstrates that the alternative scheme presented here fits within the traditional TDI framework for space based GW detectors with no additional scaling factors.
which demonstrates that the alternative scheme presented here fits within the conventional TDI framework for space based GW detectors with no additional scaling factors.
% In \cref{app:original}, we derive the original OFC-based scheme in our experimental setup as the combination $\Delta^{\tau_1}_b$ and clarify its relation to the alternative scheme presented in this paper.
For comparison, this experiment is formulated based on the original OFC-based scheme in \cref{app:original}. This is primarily to clarify the connection between the concept of the original and the proposed scheme, instead of supporting experimental results below.

\subsection{Results}\label{sub:result}
\begin{figure}[h]
    \centering
    \includegraphics[width=8.6cm]{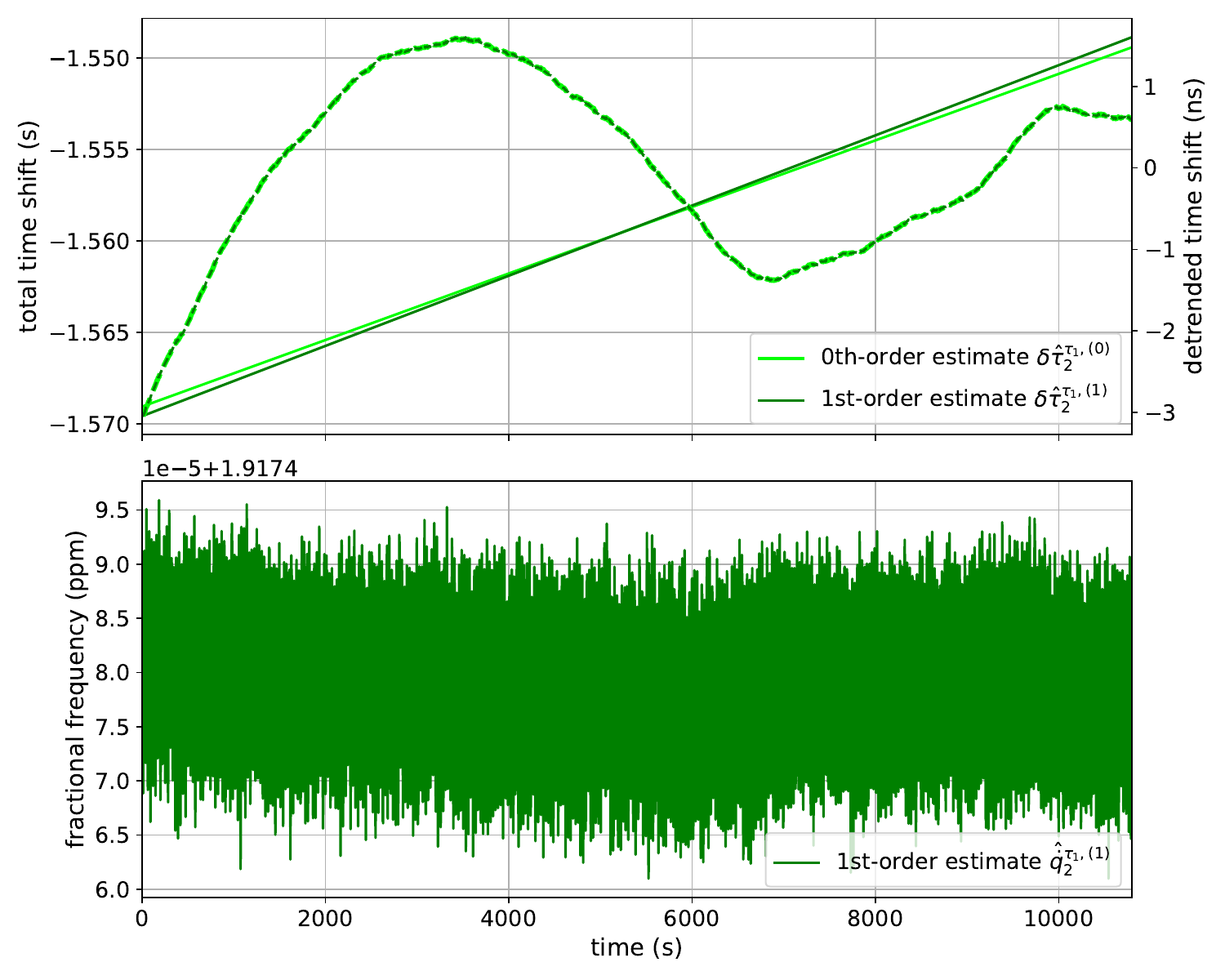}
    \caption{
    Clock difference as measured via the carrier-carrier beatnote: bottom plot shows $\hat{\dot{q}}^{\tau_1,(1)}_2$ in \cref{eq:bardotq2_tau1}; top shows $\delta\hat{\tau}_2^{\tau_1,(0)}$ in light green and $\delta\hat{\tau}_2^{\tau_1,(1)}$ in \cref{eq:deltauj_taui} in dark green.
    In the top panel, the total and detrended timer deviations are shown in solid with the left axis and dashed with the right axis, respectively.
    }
    \label{fig:clock_t}
\end{figure}

As modeled in \cref{sub:pseudorange}, the key point of our approach lies in the ability to derive the time derivative of the pseudorange from the measured carrier-carrier beatnote frequencies.
\cref{fig:clock_t} shows the timer deviation extracted from the carrier-carrier beatnote via \cref{eq:bardotq2_tau1}.
The bottom panel shows $\hat{\dot{q}}^{\tau_1,(1)}_2$, while the top panel compares the zeroth- and first-order timer deviations, $\delta\hat{\tau}_2^{\tau_1,(0)}$ and $\delta\hat{\tau}_2^{\tau_1,(1)}$, in light and dark green, respectively.
Regarding the first-order estimate, the mean value of the measured fractional frequency offset is around \SI{2.00128}{ppm}.
Against the first-order estimate, the zeroth-order estimate in light green gives a fractional frequency shift around \SI{99.4}{ppb}, which arises from the \SI{28}{\mega\Hz} difference between the primary and secondary lasers in OMS\,2.
As described in \cref{app:resolve_n}, we fit the OFC\,1 mode number $n_1$, the OFC mode number difference $n_2-n_1$, and the start time offset $T_{0,2}^{\tau_1}$ using the TDIR-like processing scheme by optimizing the fit parameters to minimize the residual noise power in the final combination $\Delta^{\tau_1,(m)}_a$.
As a result, we can derive $\hat{n}_1$, $\hat{n}_2$, and $\hat{T}_{0,2}^{\tau_1}$.
$\hat{T}_{0,2}^{\tau_1}$ was estimated to be around \SI{-1.56957818296}{\second}.

The estimated OFC parameters are summarized in \cref{tab:clock_params}.
We derive the OFC repetition rates via \cref{eq:fOFC_i2} from the estimated OFC mode numbers and other parameters, such as the nominal iodine absorption frequencies and the OFC feedback loop set points.
Notably, the derived repetition rates were compared using a separate diagnostic measurement based on an external rubidium (Rb) reference.
The estimated repetition rates agree with those measured by the Rb reference within our knowledge error of the frequencies of the iodine-stabilized lasers, namely $\pm$\SI{3}{\milli\Hz} out of \SI{100}{\mega\Hz}.
This ensures that we can identify the true OFC mode numbers exactly, as the repetition rate in our system changes by around \SI{35.5}{\Hz} with each change in the OFC mode number by 1.

\begin{table}[b]
    \caption{\label{tab:clock_params}
    Estimated OFC mode numbers and comparison of the resulting repetition rates as measured by an external Rb reference.
    $f_\mrm{rep,i}$ is derived via \cref{eq:fOFC_i2} from the estimated OFC mode numbers, the ideal iodine absorption frequency, the \SI{20}{\mega\Hz} shift by the AOM, and the OFC feedback set points.
    The \SI{3}{\milli\Hz} error of $f_\mrm{rep,i}$ is calculated from $3\sigma$ of a separate long-term measurement of the beatnote between the iodine lasers over \SI{45}{hours}, which corresponds to the accuracy of our knowledge about the iodine absorption line.
    }
    \begin{ruledtabular}
    \begin{tabular}{ll}
    Parameter & Value\\
    \hline
    $\hat{n}_1$ & 2816203
    \\
    $\hat{n}_2$ & 2816198
    \\
    $f_\mrm{rep,1}$ from $\hat{n}_1$ and I$_2$ & \SI{100003380.780}{\Hz} $\pm \SI{3.0}{\milli\Hz}$
    \\
    $f_\mrm{rep,1}$ from Rb reference & \SI{100003380.778}{\Hz} $\pm$ \SI{0.2}{\milli\Hz}
    \\
    $f_\mrm{rep,2}$ from $\hat{n}_2$ and I$_2$ & \SI{100003572.534}{\Hz} $\pm \SI{3.0}{\milli\Hz}$
    \\
    $f_\mrm{rep,2}$ from Rb reference & \SI{100003572.532}{\Hz} $\pm$ \SI{0.2}{\milli\Hz}
    \\
    \end{tabular}
    \end{ruledtabular}
\end{table}

\begin{figure}
    \centering
    \includegraphics[width=8.6cm]{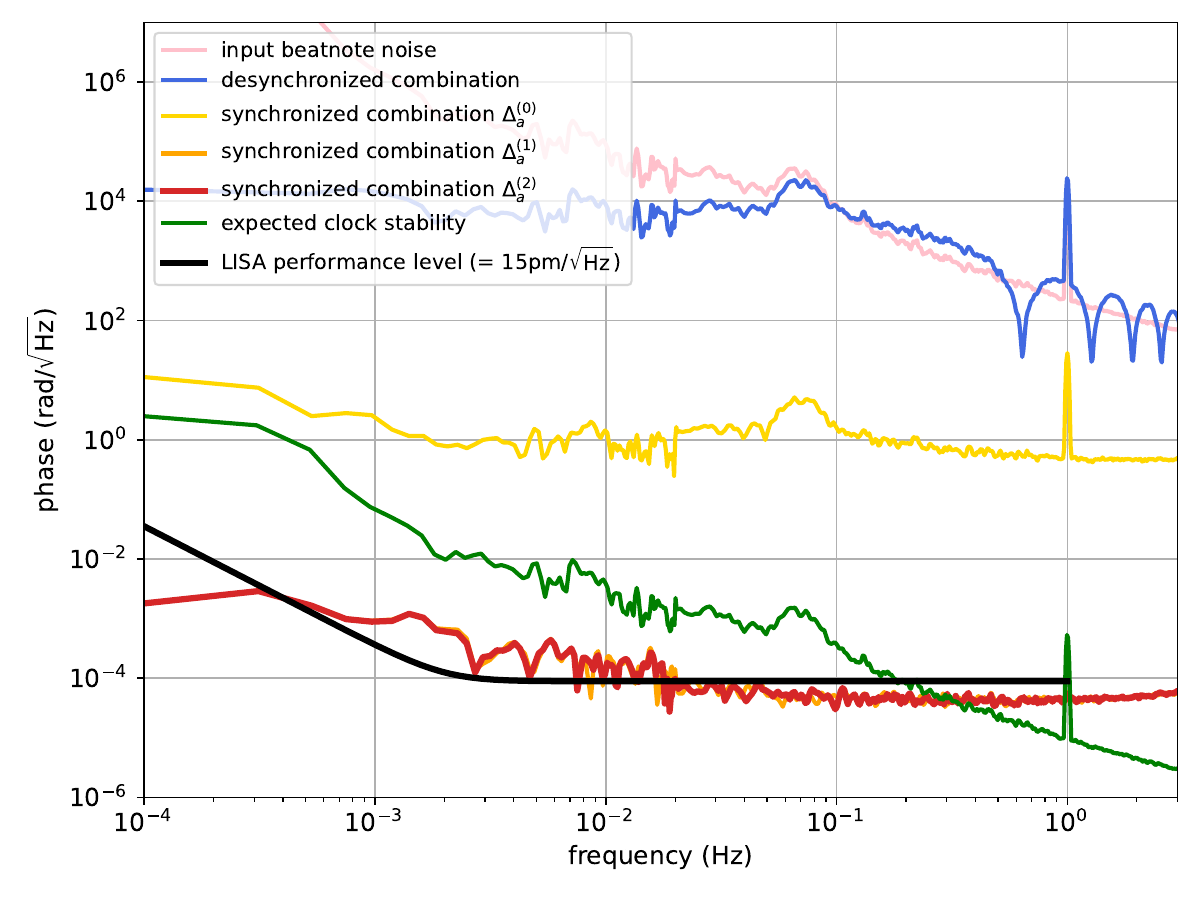}
    \caption{
    Experimental results of the noise-free combinations via our OFC-based synchronization scheme.
    The tone at \SI{1.0}{\Hz} is intentionally injected for the calibration of laser noise suppression performance.
    Pink: the input optical beatnote phase noise as measured by phasemeter 1;
    Blue: the signal combination without synchronization;
    Green: the expected in-band clock jitter after the optical-to-electrical downconversion;
    Yellow, orange, and red: the noise-free signal combination $\Delta^{\tau_1}_a$ in \cref{eq:Delta_taui_a} with iteration orders of 0, 1, and 2, respectively.
    Black: the LISA performance level, $\SI{15}{\pico\meter\prtHz}\cdot\sqrt{1 + \left(\frac{\SI{2}{\milli\Hz}}{f}\right)^4}$.
    }
    \label{fig:suppression_asd}
\end{figure}

In \cref{fig:suppression_asd}, we show the synchronization performance between the two OMSs.
The heterodyne beatnote (pink) measured by the individual phasemeters results in blue, which is the simple difference between the two without synchronization.
This is significantly affected by the start offset $T_{0,2}^{\tau_1}$ and the slow differential drift between the clocks.
Yellow, orange, and red are the noise-free combination $\Delta^{\tau_1, (m)}_a$ in \cref{eq:Delta_taui_a} with the zeroth-, first-, and second-order estimates of the timer deviation, respectively.
The zeroth-order combination is still highly affected by the inaccurate timer deviation (i.e., light green in \cref{fig:clock_t}).
By increasing the iteration order, the performance quickly converges and reaches the LISA performance level at \SI{15}{\pico\meter\prtHz}~\cite{LisaRed}; the first- and second-order combinations do not make a significant difference at the noise level.
The combination performance in red is below the in-band clock jitter (green) below \SI{0.3}{\Hz}.
This means that in addition to the synchronization of the time stamps, we can remove the stochastic jitter, as targeted by our scheme.
However, above \SI{0.3}{\Hz}, the numerical error and the aliased noise are noisier than the coupled in-band clock jitter, as discussed below; therefore, the jitter suppression is not demonstrated in the high-frequency end.
From the suppression of the injected \SI{1.0}{\Hz} tone, the synchronization accuracy is estimated to be better than \SI{0.47}{\nano\second}, which achieves the \SI{3.3}{\nano\second} mark required by LISA.

\begin{figure}
    \centering
    \includegraphics[width=8.6cm]{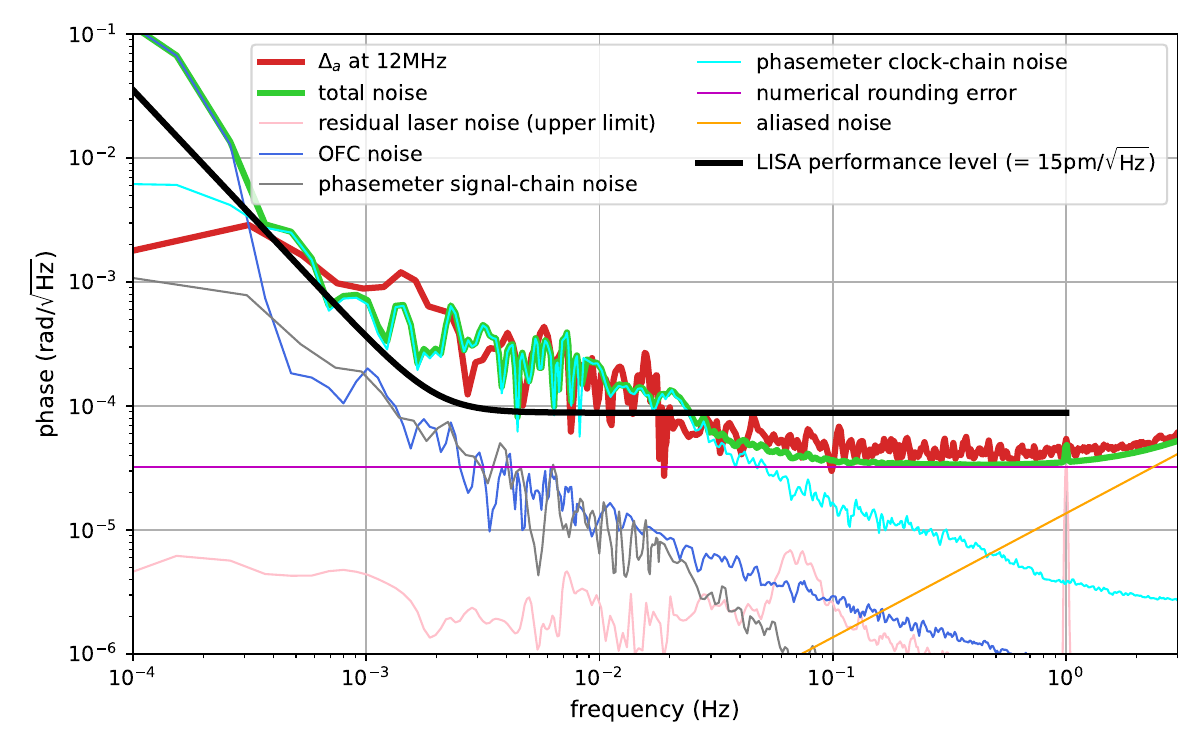}
    \caption{
    Noise budget of the signal combination $\Delta^{\tau_1,(2)}_a$.
    Red and black are identical to those in \cref{fig:suppression_asd}.
    Green is the total noise, composed of the OFC noise including peripheral RF electronics (blue), the phasemeter signal chain noise (grey), the phasemeter clock chain noise (cyan), the residual laser noise (pink), the numerical rounding error (magenta), and the aliased noise (orange).
    }
    \label{fig:noise_budget}
\end{figure}

We provide the noise budget in \cref{fig:noise_budget}.
The total noise (green) is dominated by a numerical rounding error (magenta) from \SI{10}{\milli\Hz} to \SI{1}{\Hz}.
As shown in \cref{eq:Delta_taui_a}, the combination is defined by the total phases, which increase rapidly due to the \SI{12}{\mega\Hz} heterodyne frequency.
This is not a problem if we analyze the signals in units of frequency instead of phase as is currently planned for LISA ~\cite{Bayle2021}.
Nevertheless, we employ the total phase after discovering that the frequency output of the commercial phasemeters performed worse than the phase output, and in fact worse than \SI{15}{\pico\meter\prtHz} performance requirement for LISA.
At lower frequencies, the clock chain of the commercial phasemeter is dominant.
This is expected to be further suppressed by implementing a pilot-tone correction, which enables the transformation of the in-band timing reference from the system clock to the pilot-tone signal~\cite{Xu2025,YamamotoPhD}.
At higher frequencies above \SI{1}{\Hz}, the aliased beatnote phase noise is a significant contribution.
In fact, the aliased input noise significantly affects our performance; hence, to mitigate its influence in our measurement band, we set the phasemeters to sample the data at a much higher rate, around \SI{2.4}{\kilo\Hz}, and decimated the high-rate data to around \SI{10}{\Hz} with aggressive anti-aliasing filters in postprocessing.

\begin{figure}
    \centering
    \includegraphics[width=8.6cm]{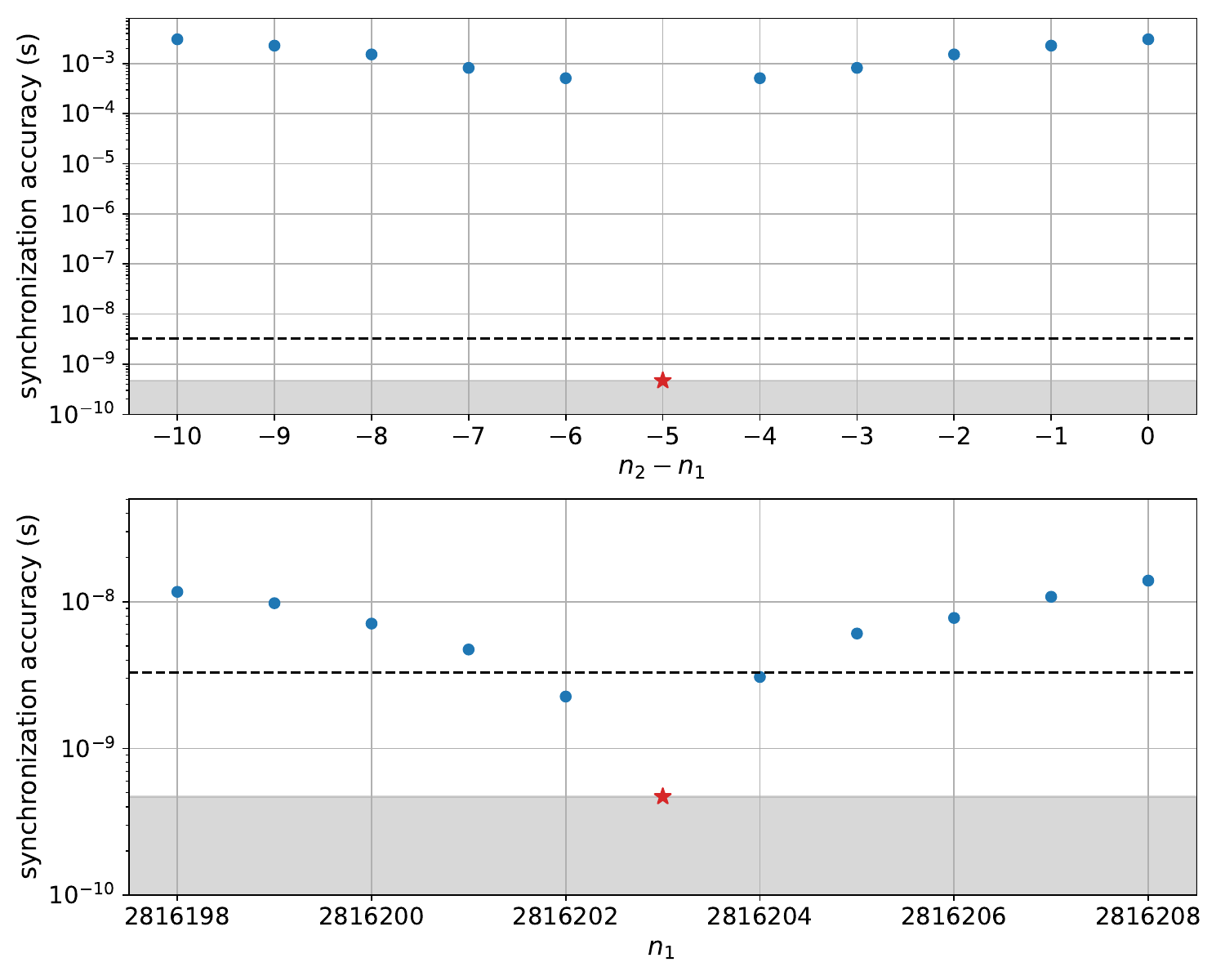}
    \caption{
    Influence of wrong OFC mode numbers on synchronization accuracy.
    Top: accuracies with the error in the difference between the mode numbers $n_2-n_1$ with $n_1$ fixed to the true value 2816203.
    Bottom: accuracies with the error in the common offset under $n_2-n_1=-5$.
    Red star: accuracies with the true mode numbers, corresponding to red in \cref{fig:suppression_asd,fig:noise_budget};
    Blue dot: accuracies with wrong mode numbers;
    Dashed black: the \SI{3.3}{\nano\second} mark;
    Grey area: accuracy below the sensitivity of the experiment.
    }
    \label{fig:nscan}
\end{figure}

Lastly, we present how the synchronization error accumulates with the error in the estimation of the OFC mode numbers.
For this purpose, in postprocessing, we fix the OFC mode numbers to an arbitrary pair of wrong values near the true integers ($n_1=2816203$ and $n_2=2816198$) and compute the synchronization performance from the residual tone at \SI{1}{\Hz} in the noise-free combination.
\cref{fig:nscan} shows the results.
The top panel shows the scan of the difference in the mode numbers, namely $n_2-n_1$, with $n_1$ fixed at the true value $2816203$.
As discussed in \cref{app:resolve_n}, the performance of the pseudorange estimation is extremely sensitive to the difference in the mode numbers: an error of 1 results in a submillisecond synchronization error, which is significant compared to the target nanosecond synchronization.
The bottom panel shows the scan of the common offset for $n_1$ and $n_2$ under $n_2-n_1=-5$.
As discussed in \cref{app:resolve_n}, the performance is much less sensitive to the common offset.
However, an error of 2 exceeds \SI{3.3}{\nano\second}.
Furthermore, the synchronization error is expected to accumulate over the measurement time, and the measurement presented here is roughly \SI{3}{hours} long.
In the end, it is necessary to identify the exact OFC mode numbers, especially for a longer measurement, which our experiment achieved via TDIR-like processing.
Once the OFC mode numbers are identified, they remain constant as long as the OFC feedback loops stay locked.
% ==================================================

\section{Conclusion}\label{sec:conclusion}
% In this paper, we present an alternative approach to time-delay interferometry using an OFC, leveraging the stochastic jitter suppression algorithm in \cite{Tinto2015,Tan2022}.
% In this paper, we present an alternative approach to TDI using an OFC, which has been developed in \cite{Tinto2015,Tan2022}.
Previous research~\cite{Tinto2015,Tan2022} established the modified TDI frameworks using OFCs.
These efforts focused on how the in-band laser and clock noises appear in the interferometric phase measurements when OFC's are incorporated.

In contrast, the alternative approach presented in this paper utilizes the carrier-carrier beatnotes to extract the time derivative of the pseudoranges, which is enabled by the coherence between the beam carriers and the OFC-derived onboard clock signals.
Compared with the EOM-based metrology adopted by LISA, this approach can be conceptually interpreted as the replacement of the clock sideband-sideband beatnote with the carrier-carrier beatnote after proper processing in \cref{eq:bardotd_leftright_taui}. %treatment.
% Hence, no change in the classical TDI formulation is required.
Hence, this approach preserves the conventional TDI formalism.
Furthermore, the time derivative of the pseudoranges includes not only the in-band stochastic jitter but also the clock frequency offset, the Doppler effect, and their slow drift.
Therefore, this study is expected to help fully leverage the technological advantages of incorporating OFCs in space-based GW detectors.
The two approaches are compared in~\cref{tab:comparison}.

\begin{table}[b]
    \caption{\label{tab:comparison}
    Comparison of the original and the proposed schemes based on OFCs in terms of noise dynamics to be suppressed, TDI formalism, and treatment for OFC mode numbers.
    }
    \begin{ruledtabular}
    \begin{tabular}{ccc}
     & Original  & Proposed \\
    \hline
    Dynamics & stochastic & stochastic + drift + offset \\
    TDI & modified & conventional \\
    OFC mode numbers $n_i$ & no need & need estimated \\
    \end{tabular}
    \end{ruledtabular}
\end{table}

To extract the time derivative of the pseudoranges from the carrier-carrier beatnotes, we need to identify the OFC mode numbers, as formulated in \cref{sub:pseudorange}.
In the experiment in \cref{sub:result}, we achieve this by fitting the comb mode integers such that the residual peak of the injected laser frequency tone at \SI{1.0}{\Hz} is minimized.
Once identified, this process does not have to be repeated unless the OFC feedback loops are broken.
We successfully synchronize the two separate phase measurement systems by an accuracy better than \SI{0.47}{\nano\second}, which achieves the \SI{3.3}{\nano\second} performance requirement of LISA.
The experiment also demonstrates the suppression of the in-band stochastic jitter down to our testbed sensitivity below the LISA performance level of \SI{15}{\pm\prtHz}.

Finally, unlike our experiment simulating only clock differences, actual LISA-like detectors include second-scale interspacecraft light travel time.
However, we expect no additional challenges caused by the long arms in terms of pseudorange derivation as the light travel time is just an additional slowly varying contribution to pseudorange, which is naturally captured by \cref{eq:bardotd_leftright_taui}.
That being said, estimating three OFC mode numbers in total (one per spacecraft) is a new complication, not existing in conventional TDI.
As done in this work, we would need them to be added to the algorithm such as TDIR at least once.
To minimize an additional computational cost of the algorithm, the fact that $n_i$ is a natural number could be efficiently exploited.

\begin{acknowledgments}
The authors thank Ira Thorpe for useful discussions.
The authors acknowledge the support from the NASA Physics of the Cosmos (PhysCOS) program. 
K. Y.’s work is supported by NASA under Award No. 80GSFC24M0006.
H.T.'s work is supported by a NASA Space Technology Graduate Research Opportunity, grant no. 80NSSC21K1277.
H.L. and C.Z.'s work is supported by the NASA Space Technology Mission Directorate  Early Career Initiative Award.
C.Z.'s work is supported in part through the Arizona NASA Space Grant Consortium, Cooperative Agreement 80NSSC25M7084.
\end{acknowledgments}

%%%%%%%% Appendix %%%%%%%%%%%%%%%%%%%%%%%%%%%%%%%%%%%
% \clearpage
\appendix
\section{Resolving OFC mode numbers $n_i$}\label{app:resolve_n}
Resolving the OFC mode number $n_i$ is required to estimate the pseudorange, as shown in \cref{eq:dotdij_taui} generally or \cref{eq:dotq2_tau1} in our experiment.
In this section, we discuss this point further.

In general, the repetition frequency $f_{\mrm{rep},i}$, which is our clock frequency, is related to the reference laser frequency $\nu_i$ via the OFC mode number $n_i$ as shown in \cref{eq:fOFC_i2}.
Let us rearrange it for $n_i$ as
\begin{align}
    n_i &= \frac{\nu_i(\tau) - f_{\mrm{OPT},i} - f_{\mrm{ceo},i}}{f_{\mrm{rep},i}(\tau)}.
    \label{eq:ni_i2}
\end{align}
Even if an absolute reference like the iodine-stabilized laser can provide the laser frequency $\nu_i$ with sufficient accuracy, we also need to know $f_{\mrm{rep},i}$ to resolve $n_i$.
The change of $f_{\mrm{rep},i}$ with $n_i$ is
\begin{align}
    \frac{\Delta f_{\mrm{rep},i}(\tau)}{\Delta n_i} &= -\frac{\nu_i(\tau) - f_{\mrm{OPT},i} - f_{\mrm{ceo},i}}{n^2_i},
    \label{eq:frep_reso}
\end{align}
which gives \SI{-35.5076060}{\Hz} with $\nu_i=\SI{281.629826}{\tera\Hz}$ (Line 1110 (R(56)32-0) $a_1$~\cite{Ye1999}), $f_{\mrm{OPT},i}=\SI{21}{\mega\Hz}$, $f_{\mrm{ceo},i}=\SI{6}{\mega\Hz}$, and $f_{\mrm{rep},i}=\SI{100}{\mega\Hz}$ (and correspondingly $n_i=2816298$).
This means that we need to know $f_{\mrm{rep},i}$ to an accuracy of \SI{0.355}{ppm} in units of fractional frequency.
Note that if we similarly estimate for $\nu_i$, we end up with almost the same required accuracy of the knowledge about $\nu_i$, but this is typically very easy to achieve with an absolute reference such as the iodine-stabilized laser.

One option would be to introduce a separate clock, such as an atomic clock, as an absolute reference more accurate than \SI{0.355}{ppm}.
However, the addition of another high performance clock in the system solely for this purpose would be a significant modification.
Therefore, it is more reasonable to perform an in-flight calibration step with the assistance of postprocessing techniques.
To gain insight into an optimal way to deal with this issue in postprocessing, let us scrutinize how the OFC mode numbers couple to the clock deviation.
To simplify this description, we limit the discussion to the clock deviation in our experiment $\dot{q}^{\tau_1}_2$
However, the discussion should be applicable to the pseudorange $\dot{d}^{\tau_i}_{ij}$ that includes the interspacecraft light travel time.

% Let us investigate how the OFC mode numbers couple to the clock deviation.
We substitute the following into \cref{eq:dotq2_tau1},
\begin{align}
    n_2 &= n_1 + a,
    \label{eq:n2_n1_a}
\end{align}
which means that $a$ is the difference between the two OFC mode numbers.
Under $f_{\mrm{OPT},12}=f_{\mrm{ceo},12}=0$, the derivative of $\dot{q}^{\tau_1}_2$ by $a$ results in
\begin{align}
    \frac{d\dot{q}^{\tau_1}_2}{da} &= -\frac{n_1}{(n_1+a)^2}\cdot \left(1+\frac{\nu^{\tau_1}_{12}(\tau)} {\nu_c}\right) \approx -\frac{1}{n_1}.
    \label{eq:ddotq_da}
\end{align}
As expected from the analysis on $f_{\mrm{rep},i}$ above, this also results in $d\dot{q}^{\tau_1}_2 \approx 3.55\cdot 10^{-7}$ when $da=1$ and $n_1 \approx 2816000$.
This means that if our estimation of the OFC mode numbers has an error of $1$ in their difference, we get \SI{0.355}{ppm} error in terms of the clock difference, which in turn amounts to a timing error of \SI{3.55}{\milli\second} after \SI{10000}{\second}.
This is a huge timing error; hence, the clock deviation $\dot{q}^{\tau_1}_2$ is highly sensitive to the error of the integer difference.

On the other hand, if we assume a common offset $b$ (that is, $n_1\rightarrow n_1+b$, $n_2\rightarrow n_2+b$), we get
\begin{align}
    \frac{d\dot{q}^{\tau_1}_2}{db} &= \frac{n_2-n_1}{(n_2+b)^2}\left(1+\frac{\nu^{\tau_1}_{12}(\tau)} {\nu_c}\right)
    \nonumber\\
    &\approx \frac{n_2-n_1}{n_2^2}
    \nonumber\\
    &\approx \frac{n_2-n_1}{n_2}\cdot \frac{d\dot{q}^{\tau_1}_2}{da},
    \label{eq:ddotq_db}
\end{align}
where $\frac{n_2-n_1}{n_2}\sim 10^{-3}-10^{-5}$.
Hence, the clock deviation $\dot{q}^{\tau_1}_2$ is much less sensitive to the common offset $b$ as expected.
However, note that a common offset estimation error of $1$ is not negligible in synchronizing the separate phase measurements to an accuracy better than \SI{3.3}{\nano\second}, or equivalently, \SI{1}{\meter}.
This is because the timing error due to the estimation error of the clock offset $\delta$ accumulates over time $T$ as $\delta\cdot T$, which can reach \SI{3.3}{\nano\second} after only \SI{1000}{\second} even if $\frac{n_2-n_1}{n_2} = 10^{-5}$ and $n_2\approx 2816000$.

Considering \cref{eq:ddotq_da,eq:ddotq_db} and the associated investigations, it is essential to derive the exact OFC mode numbers.
Hence, we can perform an in-flight calibration run, e.g., TDIR to derive those OFC mode numbers, which means optimizing the integers so that the residual noise power in the TDI combinations is minimized.
Because the clock deviation is much more sensitive to the difference rather than the common offset, we can use the mode numbers' difference (= $a$) and common offset (= $b$) as fitting parameters, instead of the individual OFC mode numbers, which can ease the optimization.

Importantly, the OFC mode numbers never change as long as the OFC feedback loops stay locked.
If the estimation protocol does not identify the exact integers and has an error in the offset $\hat{b}$ (it is very unlikely that an error will be given for the difference $\hat{a}$), the synchronization error will slowly increase over time.
However, this can be easily adjusted by adding a few integers to the estimate $\hat{n}_i$.

\section{Original scheme in our framework}\label{app:original}
To clarify the relation between the original scheme presented in ~\cite{Tinto2015} and the alternative scheme in this paper, let us write the original in the framework of our experiment in \cref{sec:theory}.
For conciseness, we assume an even simpler setup: OMS\,2 in \cref{fig:setup} also has a single unprimed laser, to which OFC\,2 is locked.
Hence, there are only the main unprimed lasers in both OMS\,1 and OMS\,2, and they directly interfere to generate a heterodyne signal tracked by the phasemeters.

Since the scheme is optimized to suppress the stochastic term of the laser and clock noise, we decompose the clock deviation $\dot{q}^{\tau_1}_2$ in \cref{eq:dotq2_tau1} into the deterministic component $\dot{q}^{\tau_1,o}_2$ and the stochastic component $\dot{q}^{\tau_1,\epsilon}_2$ as
\begin{align}
    \dot{q}^{\tau_1,o}_2(\tau) &= \frac{\hat{n}_1}{\hat{n}_2}\left(1 - \frac{O^{\tau_1}_{12}(\tau) - f_{\mrm{OPT},12} - f_{\mrm{ceo},12}}{\nu_c}\right) - 1,
    \label{eq:qijo}\\
     \dot{q}^{\tau_1,\epsilon}_2(\tau) &= \frac{\hat{n}_1}{\hat{n}_2}\cdot\frac{\dot{p}^{\tau_1}_{12}(\tau)}{\nu_c},
    \label{eq:qije}
\end{align}
where $O^{\tau_1}_{12}=O^{\tau_1}_{2}-O^{\tau_1}_{1}$ and $\dot{p}^{\tau_1}_{12}=\dot{p}^{\tau_1}_{2}-\dot{p}^{\tau_1}_{1}$.
We assume $f_{\mrm{OPT},12} = f_{\mrm{ceo},12} = 0$ below for simplicity.

Similarly, we decompose the general formula in \cref{eq:phi_i2m} to represent the time frame transformation of the measured beatnote phase between the two onboard clocks:
\begin{align}
    \phi^{\tau_2}_{12}(\tau) &= \phi^{\tau_1}_{12}\left(\tau_1^{\tau_2}(\tau)\right)
    \nonumber\\
    &\approx \phi^{\tau_1}_{12}\left(\tau_1^{\tau_2,o}(\tau)\right) -\frac{\nu^{\tau_1}_{12}\left(\tau_1^{\tau_2,o}(\tau)\right)}{1 + \dot{q}^{\tau_1,o}_2\left(\tau_1^{\tau_2,o}(\tau)\right)}\cdot q^{\tau_1,\epsilon}_2\left(\tau_1^{\tau_2,o}(\tau)\right),
    \label{eq:phi2_12}
\end{align}
which is the Taylor expansion of the phase around the deterministic part of clock 1 against clock 2, namely $\tau_1^{\tau_2,o}$.

The time stamps can be synchronized by shifting $\phi^{\tau_2}_{12}$ by the estimate of the deterministic part of the timer deviation $\delta\tau_1^{\tau_2,o}$, which is, same as \cref{eq:deltauj_taui}, the sum of a fitted start time offset $\hat{T}_{0,2}^{\tau_1}$ and the integration of $\dot{q}^{\tau_1,o}_2$ in \cref{eq:qijo}.
Regarding the stochastic component, if we substitute the integration of $\dot{q}^{\tau_1,\epsilon}_2$ in \cref{eq:qije} into \cref{eq:phi2_12}, we get the following:
\begin{align}
    \phi^{\tau_2}_{12}\left(\tau + \delta\tau_1^{\tau_2,o}(\tau)\right) &= \phi^{\tau_1}_{12}\left(\tau\right) - \frac{\nu^{\tau_1}_{12}\left(\tau\right)}{1 + \dot{q}^{\tau_1,o}_2\left(\tau\right)}\cdot q^{\tau_1,\epsilon}_2\left(\tau\right)
    \nonumber\\
    &= \phi^{\tau_1}_{12}\left(\tau\right) - \frac{\nu^{\tau_1}_{12}(\tau)}{\nu_c + O^{\tau_1}_{12}(\tau)}\cdot p^{\tau_1}_{12}(\tau)
    \nonumber\\
    &\approx \int^\tau_0 O^{\tau_1}_{12}(\tau') d\tau' + \frac{\nu_c}{\nu_c + O^{\tau_1}_{12}(\tau)}\cdot p^{\tau_1}_{12}(\tau),
    \label{eq:phi_tau2_oshift}
\end{align}
where we use $\nu^{\tau_1}_{12}(\tau) = O^{\tau_1}_{12}(\tau) + \dot{p}^{\tau_1}_{12}(\tau)$ and neglect the second-order term of the stochastic component $\dot{p}^{\tau_1}_{12}$ in the last line.

Finally, the stochastic jitter can be suppressed by properly scaling the phase signal:
\begin{align}
    \Delta^{\tau_1}_b(\tau) &= \phi^{\tau_1}_{12}(\tau) - \left(1+\frac{O^{\tau_1}_{12}(\tau)}{\nu_c}\right)\cdot\phi^{\tau_2}_{12}\left(\tau + \delta\tau_1^{\tau_2,o}(\tau)\right)
    \nonumber\\
    &= -\frac{O^{\tau_1}_{12}(\tau)}{\nu_c}\int^\tau_0 O^{\tau_1}_{12}(\tau') d\tau'.
    \label{eq:Delta_taui_b}
\end{align}
This combination leaves a deterministic residual due to the scaling factor.
Its existence in the actual TDI algorithm and impact may need to be further investigated.

The main advantage of this scheme over the alternative scheme presented in this paper is that it does not require the derivation of the OFC mode numbers $n_i$ if we neglect the clock drift.
Once we consider the clock drift and estimate it from the carrier-carrier beatnotes, we encounter the same challenge as the alternative approach, which necessitates the derivation of the OFC mode numbers.

\bibliography{main}

\end{document}